\UseRawInputEncoding
\documentclass[%
 reprint, %%%%%%%%%%%%%%%%%%%%%%%%%%%%%%%%%%%%%%%%%%%%%%%%%%%%%%%%%%%%%%%%%%%%%%%%%%%%%%%%%%%%%%%%%%%%%%%%%%%%
 superscriptaddress,
 amsmath,amssymb,
 aps,
prb,
]{revtex4-2}

\usepackage[pagewise,modulo]{lineno}
\usepackage{xcolor}
\usepackage{graphicx}% Include figure files
\usepackage{dcolumn}% Align table columns on decimal point
\usepackage{bm}% bold math
\usepackage[hidelinks,colorlinks=true,linkcolor=blue,urlcolor=blue,citecolor=blue]{hyperref}
\makeatletter
\newcommand{\mycaption}[2][]{%
  \begingroup%
  \renewcommand{\figurename}{\textbf{Figure}}
  \renewcommand{\@caption@fignum@sep}{. }%
  \renewcommand{\fnum@figure}{{\normalfont\bfseries \figurename~\thefigure}}
  \caption[#1]{#2}%
  \endgroup%
}
\makeatother

\makeatletter
\def\ignorecitefornumbering#1{%
     \begingroup
         \@fileswfalse
         #1%                     % do \cite comand
    \endgroup
}
\makeatother

\begin{document}

%\preprint{APS/123-QED}

\title{Enhanced Ferromagnetism in Monolayer Cr$_2$Te$_3$ via Topological Insulator Coupling}
\author{Yunbo Ou}
    %\email{ybou@mit.edu}
    %\thanks{These authors contributed equally}
    \affiliation{Francis Bitter Magnet Laboratory, Plasma Science and Fusion Center, Massachusetts Institute of Technology, Cambridge, Massachusetts 02139, USA}
\author{Murod Mirzhalilov}
    \affiliation{Department of Physics, The Ohio State University, Columbus, Ohio 43210, USA}
\author{Norbert M. Nemes}
    \affiliation{GFMC, Departamento F\'{i}sica de Materiales. Facultad de Ciencias F\'{i}sicas, Universidad Complutense de Madrid, 28040, Madrid, Spain}
\author{Jose L. Martinez}    
    \affiliation{Instituto de Ciencia de Materiales de Madrid ICMM-CSIC, Calle Sor Juana In\'{e}s de la Cruz, 3, Cantoblanco, Madrid 28049, Spain}
\author{Mirko Rocci}
    \affiliation{Francis Bitter Magnet Laboratory, Plasma Science and Fusion Center, Massachusetts Institute of Technology, Cambridge, Massachusetts 02139, USA}
\author{Alexander Duong}
    \affiliation{Department of Physics, University of Ottawa, Ottawa, Ontario K1N 6N5, Canada}
\author{Austin Akey}    
    \affiliation{Center for Nanoscale Systems, Harvard University, Cambridge, Massachusetts 02138, USA}
\author{Alexandre C. Foucher}    
    \affiliation{Department of Materials Science and Engineering, Massachusetts Institute of Technology, Cambridge, Massachusetts 02139, USA}
\author{Wenbo Ge}    
    \affiliation{Department of Physics and Astronomy, Rutgers University, Piscataway, New Jersey 08854, USA}
\author{Dhavala Suri}    
    %\affiliation{Francis Bitter Magnet Laboratory, Plasma Science and Fusion Center, Massachusetts Institute of Technology, Cambridge, Massachusetts 02139, USA}
    \affiliation{Centre for Nano Science and Engineering, Indian Institute of Science, Bengaluru, Karnataka 560012, India}
\author{Yiping Wang}    
    \affiliation{Department of Physics, Boston College, Chestnut Hill, Massachusetts 02467, USA}
\author{Haile Ambaye}    
    \affiliation{Neutron Scattering Division, Neutron Sciences Directorate, Oak Ridge National Laboratory, Oak Ridge, Tennessee 37831, USA}
\author{Jong Keum}    
    \affiliation{Neutron Scattering Division, Neutron Sciences Directorate, Oak Ridge National Laboratory, Oak Ridge, Tennessee 37831, USA}
    \affiliation{Center for Nanophase Materials Sciences, Physical Science Directorate, Oak Ridge National Laboratory, Oak Ridge, Tennessee 37831, USA}
\author{Mohit Randeria} 
    \affiliation{Department of Physics, The Ohio State University, Columbus, Ohio 43210, USA}
\author{Nandini Trivedi}    
    \affiliation{Department of Physics, The Ohio State University, Columbus, Ohio 43210, USA}
\author{Kenneth S. Burch}    
    \affiliation{Department of Physics, Boston College, Chestnut Hill, Massachusetts 02467, USA}
\author{David C. Bell}    
    \affiliation{Center for Nanoscale Systems, Harvard University, Cambridge, Massachusetts 02138, USA}
    \affiliation{Harvard John A. Paulson School of Engineering and Applied Sciences, Harvard University, Cambridge, Massachusetts 02138, USA}
\author{Frances M. Ross}
    \affiliation{Department of Materials Science and Engineering, Massachusetts Institute of Technology, Cambridge, Massachusetts 02139, USA}
\author{Weida Wu}    
    \affiliation{Department of Physics and Astronomy, Rutgers University, Piscataway, New Jersey 08854, USA}
\author{Don Heiman}    
    \affiliation{Francis Bitter Magnet Laboratory, Plasma Science and Fusion Center, Massachusetts Institute of Technology, Cambridge, Massachusetts 02139, USA}
    \affiliation{Department of Physics, Northeastern University, Boston, Massachusetts 02115, USA}
\author{Valeria Lauter}    
    \affiliation{Neutron Scattering Division, Neutron Sciences Directorate, Oak Ridge National Laboratory, Oak Ridge, Tennessee 37831, USA}
\author{Jagadeesh S. Moodera}    
    \email{moodera@mit.edu}
    \affiliation{Francis Bitter Magnet Laboratory, Plasma Science and Fusion Center, Massachusetts Institute of Technology, Cambridge, Massachusetts 02139, USA}
    \affiliation{Department of Physics, Massachusetts Institute of Technology, Cambridge, Massachusetts 02139, USA}
\author{Hang Chi}
    \email{hang.chi@uottawa.ca}
    %\thanks{These authors contributed equally}
    %\affiliation{Francis Bitter Magnet Laboratory, Plasma Science and Fusion Center, Massachusetts Institute of Technology, Cambridge, Massachusetts 02139, USA}
    %\affiliation{DEVCOM Army Research Laboratory, Adelphi, Maryland 20783, USA}
    \affiliation{Department of Physics, University of Ottawa, Ottawa, Ontario K1N 6N5, Canada}
    \affiliation{School of Electrical Engineering and Computer Science, University of Ottawa, Ottawa, Ontario K1N 6N5, Canada}
    \affiliation{Nexus for Quantum Technologies, University of Ottawa, Ottawa, Ontario K1N 6N5, Canada}
\date{\today}
%\date{July 6, 2022}
%%% Abstract %%%

\begin{abstract}
Exchange-coupled interfaces are pivotal in exploiting two-dimensional (2D) ferromagnetism. Due to the extraordinary correlations among charge, spin, orbital and lattice degrees of freedom, layered magnetic transition metal chalcogenides (TMCs) bode well for exotic topological phenomena. Here we report the realization of wafer-scale Cr$_2$Te$_3$ down to monolayer (ML) on insulating SrTiO$_3$(111) and/or Al$_2$O$_3$(001) substrates using molecular beam epitaxy. Robust ferromagnetism persists in the 2D limit. In particular, the Curie temperature $T_{\rm C}$ of 2~ML Cr$_2$Te$_3$ increases from 100 K to $\sim$~120~K when proximitized to topological insulator (TI) (Bi,Sb)$_2$Te$_3$, with substantially boosted magnetization as observed via polarized neutron reflectometry. Our experiments and theory strongly indicate that the Bloembergen-Rowland interaction is likely universal underlying $T_{\rm C}$ enhancement in TI-coupled magnetic heterostructures. The topological-surface-enhanced magnetism in 2D TMC enables further exchange coupling physics and quantum hybrid studies, including paving the way to realize interface-modulated topological electronics.
\end{abstract}

\maketitle

\begin{figure*}
\includegraphics{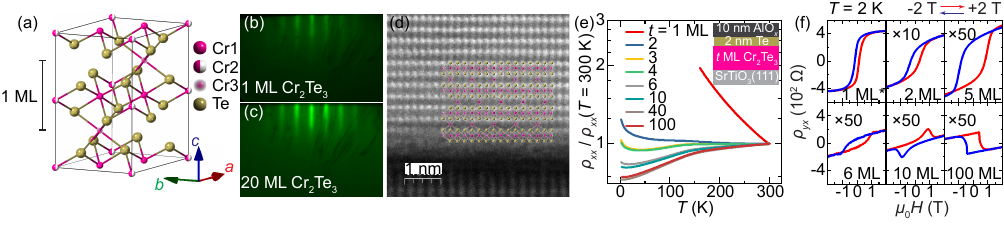}
\mycaption{\label{fig:fig1}{Structure and transport properties of Cr$_2$Te$_3$ thin films.} (a) Crystal structure of Cr$_2$Te$_3$, where three different Cr atoms are labeled as Cr1, Cr2, and Cr3. (b-c) RHEED patterns along the [100] direction of Cr$_2$Te$_3$, for (b) 1 and (c) 20 monlayer (ML) Cr$_2$Te$_3$ films grown on SrTiO$_3$(111), respectively. (d) HRSTEM HAADF image of Cr$_2$Te$_3$ on SrTiO$_3$ for the (100) plane. (e) Longitudinal electrical resistance $\rho_{xx}$ normalized to values at 300 K of Cr$_2$Te$_3$ films with thickness $t$~= 1 -- 100 ML. The inset is the schematic film structure. (f) Hall traces $\rho_{yx}$ of Cr$_2$Te$_3$ films with selected $t$~=~1~--~100~ML at 2~K. The 1~ML* data were taken from 1~ML Cr$_2$Te$_3$ with 4 QL (Bi,Sb)$_2$Te$_3$ on top. For better visibility, data for 2 ML and 5 -- 100 ML were magnified $\times 10$ and $\times 50$, respectively.}
\end{figure*}

%\noindent\textbf{Introduction.} 
\section{Introduction} 

Magnetism has been at the center of modern information technologies involving sensor, memory and logic devices \cite{2DM-Yang-2022-Nature}. Recent development of monolayer (ML) van der Waals magnetic materials \cite{2DM-Burch-2018-Nature,RN3,RN4,RN5}, such as FePS$_3$, CrI$_3$, Cr$_2$Ge$_2$Te$_6$, Fe$_3$GeTe$_2$, VSe$_2$, MnSe$_{x}$, CrSBr and AgCrP$_2$S$_6$, have revealed the feasibility of exploiting long-range order and magnetic defects for quantum sensing, information storage and processing \cite{RN6,RN7,RN8,RN9,RN10,RN11,CSB-Telford-2020,RN12,Park2024AM_Anisotropic}. Utilizing novel magnets, many appealing properties of topological nature have been investigated, including the large anomalous Hall effect (AHE), the topological Hall effect, the orbital Hall effect, spin-filtered tunneling, the topological semi-metallic and/or magnonic states \cite{RN13,RN15,Wang-NM-2019,Niu2019NC_Mixed,Chen2024NL_TopologyEngineered,Bai2024AN_Coupled}. However, wafer-size ML magnets on insulating substrates, favoring scalability for practical applications, are still much desired.

Among various magnetic transition metal chalcogenides (TMCs), Cr$_2$Te$_3$ thin films are particularly well suited for exploring magnetism towards the ML regime due to its unique crystalline, electronic and magnetic features \cite{RN_Chi_NC_2023,Cr2Te3-Zhong-2023}. The structural and chemical compatibility further allows versatile interface modulation, leveraging the salient surface properties when hybridized with Bi$_2$Te$_3$-based topological insulators (TIs) \cite{Chi_QAH_2022}. 

As shown in figure~\ref{fig:fig1}(a), Cr$_2$Te$_3$ crystallizes in a $P\bar{3}1c$ ($D_{3d}^2$, No.~163) structure with perpendicular magnetic anisotropy (PMA) along the crystallographic $c$ axis and a bulk Curie temperature $T_{\rm C}$ of $\sim$ 180 K \cite{RN21}. Along the $c$ axis, Cr$_2$Te$_3$ is composed of alternatively stacked ($i$) ferromagnetic Te-Cr1(Cr2)-Te lamellae similar to those in CrTe$_2$ \cite{RN22, CrTe2-Meng-2021, CrTe2-Zhang-2021, CrTe2-Ou-2022} and ($ii$) weakly antiferromagnetic Cr3 layers \cite{RN23} with a larger intra-layer Cr-Cr distance and often partial occupancy \cite{RN_Chi_NC_2023}. It can thus be regarded as a quasi-two-dimensional (2D) system \cite{RN24}, and we designate the Te-Cr1(Cr2)-Te-Cr3 configuration as a ML. 

Here in this Letter we report the growth of high quality Cr$_2$Te$_3$ thin films on insulating SrTiO$_3$(111) and/or Al$_2$O$_3$(001) substrates using molecular beam epitaxy (MBE). Incorporating structural, magnetic, transport and neutron measurements, we have demonstrated that the ferromagnetism in Cr$_2$Te$_3$ prevails down to 1 ML. The ordering strength in turn is favorably tunable by exchange coupling to a TI, that leads to an improved $T_{\rm C}$ \cite{RN25}. Our theoretical modeling further corroborates that topological surface states of TI are effective in stabilizing interfacial magnetic ordering at higher $T_{\rm C}$'s, serving as a \emph{general} strategy for enhancing $T_{\rm C}$ in a wide range of 2D magnetic systems including chalcogenides and oxides \cite{CGT-TI-Alegria-2014,FGT-TI-Wang-2020,LCO-TI-Zhu-2018,SRO-TI-Miao-2023}.

\begin{figure*}
\includegraphics{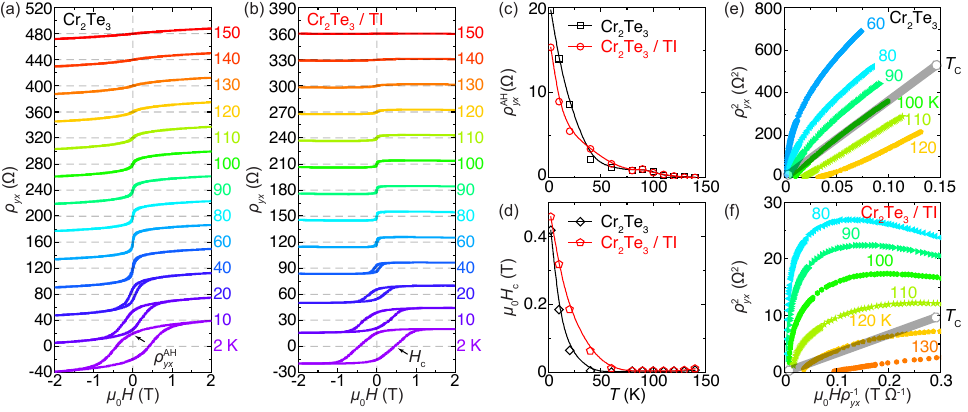}
\mycaption{\label{fig:fig2}{Topological insulator surface mediated ferromagnetism of 2 ML Cr$_2$Te$_3$.} (a-b) Magnetic field dependent anomalous Hall resistance $\rho_{yx}(H)$ of (a) 2 ML Cr$_2$Te$_3$ and (b) 2 ML Cr$_2$Te$_3$ / 4 QL (Bi,Sb)$_2$Te$_3$ heterostructure at various temperatures in the range of $T$ = 2 -- 150 K. For clarity, the curves in (a) and (b) have been vertically shifted at each temperature by 40 $\Omega$ and 30 $\Omega$, respectively. (c-d) Temperature dependence of (c) the remanent anomalous Hall resistance $\rho^{\rm AH}_{yx}$ and (d) the coercive field $H_{\rm c}$, revealing a higher Curie temperature $T_{\rm C}$ when coupled to TI. Lines are guide for the eye. (e-f) Arrott plots of $\rho^{2}_{yx}$ vs. $H/\rho_{yx}$, identifying an increase of $T_{\rm C}$ from (e) 100 K for Cr$_2$Te$_3$ to (f) $\sim$~120~K for Cr$_2$Te$_3$/TI.}
\end{figure*}

%\noindent\textbf{Structure and physical properties.} 
\section{Structure and physical properties}

Cr$_2$Te$_3$ films were grown on insulating SrTiO$_3$(111) as well as Al$_2$O$_3$(001) substrates. The streaky \emph{in situ} reflection high energy electron diffraction (RHEED) patterns shown in figures~\ref{fig:fig1}(b-c) indicate the 2D layered growth of the films down to 1 ML. The typical X-ray diffraction (XRD) patterns (supplementary figure~S1) reveal good crystallinity of Cr$_2$Te$_3$ films on SrTiO$_3$(111) substrates and are consistent with previously reported Cr$_2$Te$_3$ results \cite{RN16,RN17}. The bulk (for film with thickness $t$ = 100 ML) lattice constants were obtained as $a_{\rm XRD}$~=~6.686~($\pm$0.007)~{\AA} and $c_{\rm XRD}$~=~12.164~($\pm$0.003)~{\AA}. 

A sharp and flat interface between film and substrate is evident, as illustrated in the high-resolution scanning transmission electron microscopy (HRSTEM) high-angle annular dark-field (HAADF) image (figure~\ref{fig:fig1}(d) and supplementary figure~S2). The well-resolved region with darker contrast between Te-Cr1(Cr2)-Te trilayers corresponds to the Cr3 sites. The Raman spectra display phonon peaks that are independent of the polarization angle with respect to the crystal axes, consistent with the Cr$_2$Te$_3$ phase (supplementary figure~S3) \cite{Cr2Te3-Guillet-2023}. Measurements were performed on multiple spots across the film to verify the long-range uniformity (as also seen in the magnetic force microscopy data in supplementary figure~S3).

The thickness and temperature dependent longitudinal electrical resistance $\rho_{xx}(T)$ of Cr$_2$Te$_3$ on SrTiO$_3$(111) reveals that thick ($t$ = 100 ML) Cr$_2$Te$_3$ film is metallic at ambient temperature and displays a metal-insulator transition (MIT) like behavior at $T_{\textrm{MIT}}$ = 8 K (figure~\ref{fig:fig1}(e)). The inflection point increases at reduced $t$. Moreover, upon reducing $t$, $\rho_{xx}$ at room temperature increases in magnitude (supplementary figure~S4). Since at low temperature it becomes too insulating to measure, in the Hall experiments (figure~\ref{fig:fig1}(f)), the signal for $t$ = 1 ML was collected by probing the transport in a proximitized TI of 4 quintuple layer (QL) (Bi$_{0.23}$Sb$_{0.77}$)$_2$Te$_3$ (BST) \cite{Taylor2024MTP_Magnetotransport} on top (labeled as 1 ML*, whose excellent interface quality was demonstrated by RHEED and HRSTEM in supplementary figure S5). As shown in supplementary figure S6, 1 ML Cr$_2$Te$_3$ is believed to open an exchange gap in the proximitized 4 QL BST, leading to well established anomalous Hall signal.

The clear AHE hysteresis in the Hall resistance $\rho_{yx}(H)$ for all $t$ ranging from 1 to 100 ML, with magnetic field $H$ applied perpendicular to the $ab$ basal plane, unambiguously demonstrates the existence of ferromagnetism all the way down to 1~ML. The hysteresis in $\rho_{yx}(H)$ originates from the magnetization $M(H)$ of Cr$_2$Te$_3$. The remanent magnetization and/or Hall resistance at zero field attest to the long-range ferromagnetic order, which vanishes upon warming above~$T_{\rm C}$. 

At $T$ = 2 K, while the remanent anomalous Hall resistance $\rho^{\textrm{AH}}_{yx}$ (the zero field value of $\rho_{yx}$ after fully saturated at positive $H$) of Cr$_2$Te$_3$ is positive for the thinnest films, it changes sign to negative for $t$ $>$ 6 ML, while the ordinary Hall effect forming the linear background maintains the same positive sign for all $t$. For intermediate $t$ = 6 and 10 ML (figure~\ref{fig:fig1}(f)), $\rho_{yx}(H)$ develops a hump feature around the coercive field $H_{\textrm{c}}$. Further increase in $t$ renders bulk dominating over interface effects, recovering a square-like hysteresis in $\rho_{yx}(H)$ for 100~ML. The AHE sign reversal and hump-shaped Hall features are known to result from the interfacial strain modulated competition of channels with different Berry curvatures (that dictate the intrinsic $\rho^{\textrm{AH}}_{yx}$) and magnetic anisotropies \cite{RN_Chi_NC_2023}.

%\noindent\textbf{Monolayer magnetism tunable by TI coupling.} 
\section{Monolayer magnetism tunable by TI coupling}

The high quality of Cr$_2$Te$_3$ ultrathin films allow for in-depth investigation of intriguing exchange coupling physics. After removing the linear ordinary Hall background at high magnetic field (up to 14 T), we now turn to the rich field and temperature dependence of $\rho_{yx}$ (see supplementary figure S7). As shown in figures~\ref{fig:fig2}(a-b), the hysteresis loop visible at $T$ = 2 K gradually diminishes upon warming, for both 2 ML Cr$_2$Te$_3$ and 2 ML Cr$_2$Te$_3$/TI. The temperature dependent remanent $\rho^{\rm AH}_{yx}$ and coercivity $H_{\rm c}$ are extracted and summarized in figures~\ref{fig:fig2}(c) and (d), respectively. It is evident that the onset of non-zero $H_{\rm c}(T)$ reveals a higher $T_{\rm C}$ and stronger magnetism for the TI-proximitized film (see also supplementary figure S6). Since the $H_{\rm c}(T)$ profiles of the two samples tend to overlap at higher $T$, Arrott diagrams \cite{Arrott-1957} -- plotting $\rho^{2}_{yx}$ vs. $H/\rho_{yx}$, where the curvature of the isotherms changes at the critical point -- are used to more accurately determine the $T_{\rm C}$ = 100 K (figure~\ref{fig:fig2}(e)) and $\sim$~120~K (figure~\ref{fig:fig2}(f)) for 2 ML Cr$_2$Te$_3$ and 2 ML Cr$_2$Te$_3$/TI samples, respectively.

\begin{figure*}
\includegraphics[width=1\textwidth]{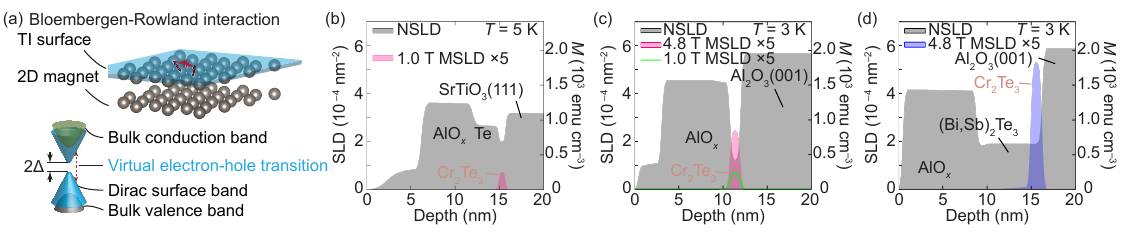}
\mycaption{\label{fig:fig3}\textbf{Polarized neutron reflectometry of Cr$_2$Te$_3$ thin films and their heterostrucutres with (Bi,Sb)$_2$Te$_3$.} (a) Enhancing 2D magnetism via the Bloembergen-Rowland interaction in proximity with a topological insulator (TI). (b-d) Depth profiles of nuclear (NSLD) and magnetic (MSLD) scattering length densities (SLD), for pristine Cr$_2$Te$_3$ monolayers (b) on SrTiO$_3$(111) substrate (measured with in-plane magnetic field $\mu_{0}H$ = 1.0 T at $T$ = 5 K), (c) on Al$_2$O$_3$(001) substrate ($\mu_{0}H$ = 1.0 T and 4.8 T at $T$ = 3 K), and (d) interfaced with (Bi,Sb)$_2$Te$_3$ TI ($\mu_{0}H$ = 4.8 T at $T$ = 3 K), respectively.}
\end{figure*}

Polarized neutron reflectometry (PNR) was invoked to further examine the ferromagnetism of Cr$_2$Te$_3$ and its coupling to TI in the monolayer regime \cite{pnr-method-3,pnrLauter-2012}. As shown in figure~\ref{fig:fig3}, to probe the TI-strengthened magnetic ordering, the depth-sensitive PNR experiments were carried out on 1~ML Cr$_2$Te$_3$ film on SrTiO$_3$(111) substrate, 2~ML Cr$_2$Te$_3$ films on Al$_2$O$_3$(001) substrate, as well as coupled to a top BST layer using Al$_2$O$_3$ substrate. The spin asymmetry (SA) ratio = $(R^{+}-R^{-})/(R^{+}+R^{-})$ was measured as a function of the wave vector transfer $Q = 4\pi\sin(\theta)/\lambda$, where $R^+$ and $R^-$ are the reflectivity for the neutron spin parallel ($+$) or antiparallel ($-$) to the applied magnetic field, respectively. As shown in supplementary figure~S8, the non-zero SA clearly reveals the emergence of robust magnetization in 2 ML samples and its enhancement when coupled to TI surfaces. By refining the PNR data, the corresponding depth profiles of the nuclear (NSLD) and magnetic (MSLD) scattering length densities (SLD) were obtained. 

As illustrated in figures~\ref{fig:fig3}(b) and (c) with $\mu_0H$ = 1.0~T, Cr$_2$Te$_3$ films manifest similar magnetization $M$ of 52 ($\pm 10$) emu~cm$^{-3}$ and 47 ($\pm 10$) emu~cm$^{-3}$ when grown on SrTiO$_3$ (measured at 5 K) and Al$_2$O$_3$ (measured at 3 K). Figures~\ref{fig:fig3}(c) and (d) reveal that, under a higher $\mu_0H$ = 4.8~T that saturates the in-plane magnetic moments, $M$ increases from 173 ($\pm 17$) emu~cm$^{-3}$ for Cr$_2$Te$_3$ to 369 ($\pm 17$) emu~cm$^{-3}$ when coupled with TI. Furthermore, the enhancement in $M$ is also consistent with the observation under $\mu_0H$ = 1.0~T that $M$ = 47 ($\pm 10$) emu cm$^{-3}$ for Cr$_2$Te$_3$ (measured at 3~K) increases to 157 ($\pm 14$) emu cm$^{-3}$ of BST-Cr$_2$Te$_3$-BST trilayers (measured at 11 K) in supplementary figure~S9. Despite the choice of the elevated $T$ = 11 K due to technical reasons when measuring the trilayer sample, $M$ for the trilayer sandwiched structure is expected to be even higher when cooled to the base $T$ = 3 K. The PNR observation attests to the effectiveness of TI-enhanced magnetism.

%\noindent\textbf{Theory on topological surface mediated coupling.} 
\section{Theory on topological surface mediated coupling}

We next aim to gain insight into the nature of the interfacial interaction between TI surface states and the local moments in a 2D magnet, and to understand the enhancement of $T_{\rm C}$ and magnetic anisotropy. Toward this end, we construct a model Hamiltonian $H=H_0+H_{\rm int}$, where $H_0$ is the kinetic energy of electrons in the 2D topological surface states of the 3D TI (BST), and $H_{\rm int}$ is the Kondo or $s$-$d$ interaction between these electrons and localized Cr moments in Cr$_2$Te$_3$ at the interface. 

The Dirac Hamiltonian $H_0=\sum_{\boldsymbol{k},\lambda\lambda'}\hat{c}^\dagger_{\boldsymbol{k}\lambda} [\hbar v_{\rm F} (\boldsymbol{k}\times\boldsymbol{\sigma}^{\lambda\lambda'})\cdot\hat{\mathbf{z}}+\Delta\sigma^{\lambda\lambda'}_z]\hat{c}_{\boldsymbol{k}\lambda'}$ describes the surface state electrons on energy scales less than the bandwidth~$W (=\hbar v_{\rm F}/a)$. Here $v_{\rm F}$ is the Fermi velocity, $\boldsymbol{k}$ is the in-plane momentum, $\boldsymbol{\sigma}$'s are the spin Pauli matrices, $\hat{\mathbf{z}}$ is the unit vector normal to the surface, $\lambda,\lambda'$ are spin labels, $\Delta$ is the exchange gap due to broken time-reversal symmetry at the interface and $a$ is the lattice constant of Cr ions. 
$H_{\rm int}=J a^2\sum_{i,\lambda\lambda'}\mathbf{S}_i\cdot\hat{c}_{i\lambda}^\dagger\boldsymbol{\sigma}^{\lambda\lambda'}\hat{c}_{i\lambda'}$ characterizes the Kondo coupling $J$ (with units of energy) between a local moment $\mathbf{S}_i$ at $\mathbf{R}_i$ and the surface state electron spin.  

Our first goal is to obtain the effective Hamiltonian
\begin{equation}\label{EqHeff}
    H_{\rm eff}=-\frac{1}{2}\sum_{ij,\alpha\beta}J_{\alpha\beta}(\mathbf{R}_{ij})S_i^{\alpha}S_j^{\beta},
\end{equation}
which describes the topological surface state-mediated exchange coupling $J_{\alpha\beta}(\mathbf{R}_{ij})$ between the classical spins $\mathbf{S}_i$.
Here $\mathbf{R}_{ij}=\mathbf{R}_i-\mathbf{R}_j$ and $\alpha,\beta$ represent $x,y,z$ spin components. Using second-order perturbation theory in $J/W$, we find $H_{\rm eff}^{(ij)}=-J^2\sum_{\alpha\beta} \chi_{\alpha\beta}(\mathbf{R}_{ij})S_{i\alpha}S_{j\beta}$, where $\chi_{\alpha\beta}(\mathbf{R}_{ij})$ is the static spin susceptibility of the surface state electrons. At zero temperature \cite{Imamura2004}, \( \chi_{\alpha\beta}(\mathbf{R}_{ij})=(1/\pi) \text{Im}\int_{-\infty}^{\mu}d\omega \text{Tr}[\sigma_\alpha G(\mathbf{R}_{ij},\omega)\sigma_\beta G(-\mathbf{R}_{ij},\omega)] \), where $\mu$ is the chemical potential, the trace is over spin, and \( G(\mathbf{R}_{ij},\omega) \) is the retarded electronic Green's function whose form is known analytically~\cite{Galitski2014,MM}. The integral above can be split into two parts: the $\int_{-\infty}^0$ term describes the contribution from the filled valence band, while the $\int_{0}^\mu$ piece arises from the Fermi surface of the partially filled conduction band. The latter leads to an oscillatory Ruderman-Kittel-Kasuya-Yosida (RKKY) coupling $\sim \cos(2k_{\rm F} R)/R^2$ for 
$k_{\rm F} R \gg 1$, where $k_{\rm F}$ is the Fermi wave vector \cite{Chang2011,Zhang2009}.

The first term is ferromagnetic and decays as $1/R^3$ for $a \leq R \ll \hbar v_{\rm F}/\Delta$ and exponentially for $R \gg \hbar v_{\rm F}/\Delta$. We find~\cite{MM} that this piece dominates over the RKKY contribution on the distance scale of $R \sim a$ when we are in the low-doping regime $k_{\rm F} a \ll 1$. Therefore, we neglect the RKKY term and set the chemical potential to zero. The resulting coupling,
 known as the Bloembergen-Rowland (BR) interaction \cite{BR1955}, describes virtual interband transitions from the filled valence band (figure~\ref{fig:fig3}a).

For simplicity, we assume that the Cr1 and Cr2 sites are the same and this results in a triangular lattice of Cr local moments. We now use the effective Hamiltonian \eqref{EqHeff} describing spins on a triangular lattice to analyze its properties using several different methods~\cite{MM}. First, we show that for small clusters of spins, the ground state is ferromagnetic in the out-of-plane direction. Next, by computing the spin-wave spectrum in the full lattice model, we found that the spectrum is gapped and of the form $\omega(q)=K+Aq^2$ near $q=0$. Here $A > 0$ is the ferromagnetic exchange stiffness while the gap at $q=0$ indicates a PMA of the form $K=f(a\Delta/\hbar v_{\rm F}) J^2S/W$. We find numerically that the dimensionless function 
$f \approx 0.15$ when $a\Delta/\hbar v_{\rm F}$ lies in the realistic range of $0-0.1$. The magnetic anisotropy arises from the anisotropic nature of the BR interaction, which in turn results from the spin-momentum locking of the surface states~\cite{Pesin2011}. The exchange and PMA stabilize an out-of-plane ferromagnetic order at the interface.

\begin{table}%[b]
\caption{\label{tab:TC} Magnetism modulated by topological insulator coupling. The Curie temperature $T_{\rm C}$ of various magnetic materials increases to $T^\prime_{\rm C}$ when interfaced with a topological insulator, displaying an enhancement $\Delta T_{\rm C}$.}
\begin{ruledtabular}
\begin{tabular}{ccccc}
Material & $T_{\rm C}$ (K) & $T^\prime_{\rm C}$ (K) & $\Delta T_{\rm C}$ (K) & Ref. \\
\hline
2 ML Cr$_2$Te$_3$ & 100 & 120 & 20 & This work\\
Cr$_2$Ge$_2$Te$_6$ & 61 & 108 & 47 & \cite{CGT-TI-Alegria-2014}\\
Fe$_3$GeTe$_2$ & 230 & 400 & 170 & \cite{FGT-TI-Wang-2020}\\
LaCoO$_3$ & 85 & 100 & 15 & \cite{LCO-TI-Zhu-2018}\\
SrRuO$_3$ & 105 & 130 & 25 & \cite{SRO-TI-Miao-2023}\\
\end{tabular}
\end{ruledtabular}
\end{table}

We next perform a mean-field calculation for the full lattice Hamiltonian \eqref{EqHeff}, which leads to \(\Delta T_{\rm C} = [{J^2S(S+1)}/{3k_{\rm B}}]\sum_{\mathbf{R}}\chi_{zz}(\mathbf{R}) \) that is proportional to the Van Vleck susceptibility. Using parameters relevant to the Cr$_2$Te$_3$/BST system: \( S=3/2 \), $a$ = 3.93~{\AA}, $v_{\rm F}=3.69\times10^{5}$ m s$^{-1}$ \cite{BST2015}, and $J\approx$ 0.1 eV \cite{Chang2011,Zhang2009}. The bandwidth $W = \hbar v_{\rm F}/a \simeq 0.6$ eV which justifies the perturbation theory in $J/W$. We note that at $T_{\rm C}$ one loses the time-reversal breaking gap $\Delta$, which can then be set to zero. We thus find a mean-field enhancement of $T_{\rm C}$ to be $\Delta T_{\rm C} \approx 50$ K arising from the BR interaction. Since mean-field theory ignores the effects of fluctuations and overestimates the transition temperature, this answer is in reasonably good agreement with our experimental observation of $T_{\rm C}$ enhancement from transport and PNR.

The theoretical framework we have developed here can be generalized to model the experimental observations of $T_{\rm C}$ increase near the interface in other TI-proximitized magnets, such as Cr$_2$Ge$_2$Te$_6$ \cite{CGT-TI-Alegria-2014}, Fe$_3$GeTe$_2$ \cite{FGT-TI-Wang-2020}, LaCoO$_3$ \cite{LCO-TI-Zhu-2018} and SrRuO$_3$ \cite{SRO-TI-Miao-2023}, as summarized in Table ~\ref{tab:TC}. In addition to parameters specific to each system, our modeling would also need to take into account that Fe$_3$GeTe$_2$ and SrRuO$_3$ are metallic magnets, distinct from the localized moments in the few ML Cr$_2$Te$_3$ treated here \cite{Cr2Te3-Zhong-2023}. We will report on this in a separate theory paper \cite{MM}.

%\noindent\textbf{Conclusion.} 
\section{Conclusion}

In summary, the observation of 2D ferromagnetism in monolayer Cr$_2$Te$_3$ MBE films on insulating SrTiO$_3$(111) and Al$_2$O$_3$(001) substrates has uncovered a novel atomically thin ferromagnetic candidate capable of strong proximity exchange coupling with other van der Waals quantum materials such as topological insulators for tunable long range order. In particular, via the Bloembergen-Rowland interaction prevalent when coupling topological insulators to 2D magnets, our comprehensive transport measurements, PNR experiments and theoretical modeling identify topological surface state proximity as a highly versatile pathway enhancing $T_{\rm C}$ in a wide family of magnets. The interface modulated 2D ferromagnetism in Cr$_2$Te$_3$ and related magnetic transition metal chalcogenide films bodes well for future magnetic topological device design towards all-van der Waals \cite{RN27} and molecular \cite{Molecular-Raman-2013} spintronics.

\section*{Methods}
\textbf{Sample growth.} The samples were grown in an ultrahigh vacuum (UHV) MBE system with a base pressure $< 3.7 \times 10^{-10}$ Torr. The growth was monitored by a RHEED system (STAIB 20) with an electron energy of 15 keV. The SrTiO$_3$(111) substrate was rinsed in distilled water at 80~$^\circ$C for 90 minutes and then annealed in a tube furnace at 950~$^\circ$C with oxygen flow for 3 hours. Before film growth, the substrate was degassed at 500~$^\circ$C for 10 min and then heated at 600~$^\circ$C for 25 min in the MBE chamber. High purity Bi (99.999{\%}), Sb (99.9999{\%}), Te (99.999{\%}), and Cr (99.999{\%}) were co-evaporated from Knudsen cell evaporators and/or e-guns. The flux of each element was monitored by individual quartz crystal monitor during the growth. The growth of Cr$_2$Te$_3$ was conducted under Te-rich conditions at a substrate temperature of 235~$^\circ$C with a typical Te/Cr flux ratio of 10. The BST films were grown under the same Te-rich conditions at a substrate temperature of 235~$^\circ$C with a typical Te/Bi,Sb flux ratio of 10. The Bi/Sb ratio was picked to optimize the resultant chemical potential into the bulk band gap. A 2 nm Te layer and a subsequent 10 nm AlO$_x$ layer were \emph{in situ} deposited on the films at room temperature to protect the film from degradation by air exposure.\\

%\newpage

\textbf{Raman spectroscopy.} The samples for Raman spectroscopy were capped with \emph{in situ} grown 10 nm AlO$_x$. Raman scattering was performed with a custom built, low temperature microscopy setup \cite{RN28}, using a 532 nm excitation laser that has a spot size of 2 $\mu$m in diameter. The spectrometer has a 2400~g~mm$^{-1}$ grating, with an Andor CCD, providing a resolution of $\sim$~1~cm$^{-1}$. All data presented in this work were taken at room temperature, and laser heating was minimized using the Stokes to anti-Stokes ratio \cite{RN29}. Polarization dependence was accomplished by linearly polarizing the excitation laser in the sample plane and rotating the polarization direction via a $\lambda/2$ Fresnel rhomb. A second polarizer was used to analyze the scattered light, which was either parallel (XX) or perpendicular (XY) to the incoming polarization direction. Dark counts were removed by subtracting data collected with the same integration time with the laser blocked. Furthermore, a recently developed wavelet-based approach was employed to remove the effect of cosmic Rays \cite{RN30}.\\

\textbf{X-ray characterizations.} The XRD patterns were obtained using a parallel beam of Cu K$_{\alpha1}$ radiation with wavelength $\lambda$ = 0.15406~nm in a Rigaku SmartLab system. The $2\theta$ (for OOP measurement) and/or $2\theta_{\chi}$ (for IP configuration) scan angles were typically between 10$^\circ$ and 120$^\circ$ with a step size of 0.05$^\circ$. XRR measurements were performed at the Center for Nanophase Materials Sciences (CNMS), Oak Ridge National Laboratory, on a PANalytical X'Pert Pro MRD equipped with hybrid monochromator and Xe proportional counter. For the XRR measurements, the X-ray beam was generated at 45~kV/40~mA, and the X-ray beam wavelength after the hybrid mirror was $\lambda$ = 0.15406~nm (Cu K$_{\alpha1}$ radiation).\\

\textbf{Scanning transmission electron microscopy.} STEM experiments were conducted at a probe-corrected STEM (JEOL ARM) operated at an acceleration voltage of 200~kV. Samples were prepared by a Helios focused-ion beam (FEI), operated at an acceleration voltage of 30~kV for the gallium beam during lift-out and of 2~kV during polishing. Additional polishing was performed at 1~kV and 0.5~kV with a NanoMill (Fischione). At both acceleration voltages, samples were polished for 20~min on each side. For the Cr$_2$Te$_3$/TI sample, STEM images were acquired with a Themis Z G3 instrument manufactured by Thermo Fischer Scientific operated at 200 kV with a beam current of 40 pA and a convergence semi-angle of 20 mrad.\\

\textbf{Transport and magnetic measurement.} Transport measurements were performed in a Quantum Design Physical Property Measurement System (PPMS, 1.9 K, 14 T). The film was manually scratched into a Hall bar geometry. The electrodes were made by mechanically pressing fine indium pieces onto the contact areas of the film. The typical sample size was 500~$\mu$m~$\times$~200~$\mu$m. Magnetic properties were measured in a Quantum Design Magnetic Property Measurement System 3 (MPMS3, 1.8~K, 7~T). The typical sample size was 5 mm $\times$ 5 mm.\\

\textbf{Magnetic force microscopy.} The MFM experiments were carried out in a homemade cryogenic atomic force microscope (AFM) using commercial piezoresistive cantilevers (spring constant $k$ $\approx$ 3~N~m$^{-1}$, resonant frequency $f_0 \approx 42$~kHz). The homemade AFM is interfaced with a Nanonis SPM Controller using a phase-lock loop (SPECS). MFM tips were prepared by depositing a 150~nm Co film onto bare tips using electron-beam evaporation. MFM images were taken in a constant height mode with the scanning plane $\sim$~20~nm above sample surface. The MFM signal, the change of the cantilever resonant frequency ${\delta}f$, is proportional to the OOP force gradient acting on the tip, which is a second derivative of the stray field.\\

\textbf{Polarized neutron reflectometry.}  PNR is a highly penetrating depth-sensitive technique to probe the chemical and magnetic depth profiles with a resolution of 0.5~nm. The depth profiles of the NSLD and MSLD correspond to the depth profile of the chemical and IP magnetization vector distributions on the atomic scale, respectively \cite{pnr-method-1,pnr-method-2,pnr-method-3}. Based on these neutron scattering merits, PNR serves as a powerful technique to simultaneously and nondestructively characterize chemical and magnetic nature of buried interfaces \cite{pnrLauter-2012}. PNR experiments were performed on the Magnetism Reflectometer at the Spallation Neutron Source at Oak Ridge National Laboratory \cite{pnrValeria, pnrTong, pnrSyro, pnrJiang}, using neutrons with wavelengths $\lambda$ in a band of 0.2--0.8~nm and a high polarization of 98.5--99{\%}. Measurements were conducted in a closed cycle refrigerator (Advanced Research System) equipped with a 1.15~T electromagnet (Bruker) and in a top-loading closed cycle refrigerator combined with a 5~T cryomagnet (Cryomagnetics). Using the time-of-flight method, a collimated polychromatic beam of polarized neutrons with the wavelength band $\delta\lambda$ impinged on the film at a grazing angle $\theta$, interacting with atomic nuclei and the spins of unpaired electrons. The reflected intensity $R^+$ and $R^-$ were measured as a function of momentum transfer, $Q = 4\pi\sin(\theta)/\lambda$, with the neutron spin parallel ($+$) or antiparallel ($-$), respectively, to the applied field. To separate the nuclear from the magnetic scattering, the spin asymmetry ratio SA = $(R^{+}-R^{-})/(R^{+} + R^{-})$ was calculated, with SA = 0 designating no magnetic moment in the system. Being electrically neutral, spin-polarized neutrons penetrate the entire multilayer structures and probe magnetic and structural composition of the film and buried interfaces down to the substrate.\\

\noindent\textbf{Data availability.} The data that support the findings of this study are available from the corresponding authors upon reasonable request.\\

\noindent\textbf{Acknowledgements.}
This work was supported by Army Research Office (ARO W911NF-20-2-0061), the National Science Foundation (NSF-DMR 2218550), Office of Naval Research (N00014-20-1-2306). H.C. acknowledges support of the Canada Research Chairs (CRC) Program, and the Natural Sciences and Engineering Research Council of Canada (NSERC), Discovery Grant RGPIN-2024-06497 and ALLRP 592642-2024. Y.O., D.S., J.S.M. and D.C.B. thank the Center for Integrated Quantum Materials (NSF-DMR 1231319) for financial support. This work made use of the MIT Materials Research Laboratory. D.H. thanks support from NSF-DMR 1905662 and the Air Force Office of Scientific Research Award No. FA9550-20-1-0247. N.M.N. and J.L.M. were supported by Ministerio de Ciencia e Innovacion, Spain, with grant numbers MAT2017-84496-R, PID2021-122477OB-I00, TED2021-129254B-C21 and TED2021-129254B-C22. M.R. received funding from the European Union Horizon 2020 research and innovation programme under the Marie Sklodowska-Curie grant agreement EuSuper No. 796603. The STEM work was performed in part at the Center for Nanoscale Systems (CNS), a member of the National Nanotechnology Coordinated Infrastructure Network (NNCI), which is supported by NSF Award No. 1541959. A.C.F. is supported by the MIT-IBM Watson AI Lab. This work was carried out in part through the use of MIT.nano's facilities. The MFM studies at Rutgers were supported by the Office of Basic Energy Sciences, Division of Materials Sciences and Engineering, U.S. Department of Energy under Award No. DE-SC0018153. The Raman measurements of Y.W. and K.S.B. were performed with support from the Air Force Office of Scientific Research under Award No. FA9550-24-1-011. This research used resources at the Spallation Neutron Source, a Department of Energy Office of Science User Facility operated by the Oak Ridge National Laboratory (ORNL). Neutron reflectometry measurements, beamtime proposal IPTS-24447, were carried out on the Magnetism Reflectometer at the SNS, which is sponsored by the Scientific User Facilities Division, Office of Basic Energy Sciences, DOE. XRR measurements were conducted at the Center for Nanophase Materials Sciences (CNMS) at ORNL, which is a DOE Office of Science User Facility. M.M., M.Ra. and N.T. were supported by the NSF Materials Research Science and Engineering Center Grant No. DMR-2011876. N.T. thanks K. Lee for discussions. Notice: This manuscript has been authored by UT-Battelle, LLC under Contract No. DE-AC05-00OR22725 with the U.S. Department of Energy. The United States Government retains and the publisher, by accepting the article for publication, acknowledges that the United States Government retains a non-exclusive, paid-up, irrevocable, world-wide license to publish or reproduce the published form of this manuscript, or allow others to do so, for United States Government purposes. The Department of Energy will provide public access to these results of federally sponsored research in accordance with the DOE Public Access Plan \cite{ORNL_standard_disclaimer}.\\

\noindent\textbf{Author contributions.}
%\noindent\textbf{Author contributions}
H.C. conceived the project. The samples were prepared and characterized by Y.O., H.C. and J.S.M.. N.M.N., J.L.M., M.Ro., Y.O., A.D. and H.C. performed transport measurements and analyzed results. Y.O., H.C. and D.H. carried out magnetization measurements. A.A., D.C.B., A.C.F. and F.M.R. collected and analyzed HRSTEM images. W.G. and W.W. obtained MFM images. D.S., Y.W. and K.S.B. measured Raman spectra. J.K. performed XRR measurements, V.L. and H.A. conducted PNR experiments, V.L. analyzed XRR and PNR data. M.M., M.Ra. and N.T. provided theoretical modeling and analysis. H.C. wrote the paper with input from all authors. All authors discussed the results.

% The \nocite command causes all entries in a bibliography to be printed out
% whether or not they are actually referenced in the text. This is appropriate
% for the sample file to show the different styles of references, but authors
% most likely will not want to use it.

%\nocite{*}

\bibliography{0-MS}% Produces the bibliography via BibTeX.

\end{document}

%\preprint{APS/123-QED}

%\title{Supplementary Information \linebreak 
%Strain tunable Berry curvature in quasi-two-dimensional chromium telluride}
%\author{H. Chi \textit{et al.}}    
%\date{\today}

\title{Supplementary Material \linebreak Enhanced Ferromagnetism in Monolayer Cr$_2$Te$_3$ via Topological Insulator Coupling}
\author{Yunbo Ou}
    %\email{ybou@mit.edu}
    %\thanks{These authors contributed equally}
    \affiliation{Francis Bitter Magnet Laboratory, Plasma Science and Fusion Center, Massachusetts Institute of Technology, Cambridge, Massachusetts 02139, USA}
\author{Murod Mirzhalilov}
    \affiliation{Department of Physics, The Ohio State University, Columbus, Ohio 43210, USA}
\author{Norbert M. Nemes}
    \affiliation{GFMC, Departamento F\'{i}sica de Materiales. Facultad de Ciencias F\'{i}sicas, Universidad Complutense de Madrid, 28040, Madrid, Spain}
\author{Jose L. Martinez}    
    \affiliation{Instituto de Ciencia de Materiales de Madrid ICMM-CSIC, Calle Sor Juana In\'{e}s de la Cruz, 3, Cantoblanco, Madrid 28049, Spain}
\author{Mirko Rocci}
    \affiliation{Francis Bitter Magnet Laboratory, Plasma Science and Fusion Center, Massachusetts Institute of Technology, Cambridge, Massachusetts 02139, USA}
\author{Alexander Duong}
    \affiliation{Department of Physics, University of Ottawa, Ottawa, Ontario K1N 6N5, Canada}
\author{Austin Akey}    
    \affiliation{Center for Nanoscale Systems, Harvard University, Cambridge, Massachusetts 02138, USA}
\author{Alexandre C. Foucher}    
    \affiliation{Department of Materials Science and Engineering, Massachusetts Institute of Technology, Cambridge, Massachusetts 02139, USA}
\author{Wenbo Ge}    
    \affiliation{Department of Physics and Astronomy, Rutgers University, Piscataway, New Jersey 08854, USA}
\author{Dhavala Suri}    
    %\affiliation{Francis Bitter Magnet Laboratory, Plasma Science and Fusion Center, Massachusetts Institute of Technology, Cambridge, Massachusetts 02139, USA}
    \affiliation{Centre for Nano Science and Engineering, Indian Institute of Science, Bengaluru, Karnataka 560012, India}
\author{Yiping Wang}    
    \affiliation{Department of Physics, Boston College, Chestnut Hill, Massachusetts 02467, USA}
\author{Haile Ambaye}    
    \affiliation{Neutron Scattering Division, Neutron Sciences Directorate, Oak Ridge National Laboratory, Oak Ridge, Tennessee 37831, USA}
\author{Jong Keum}    
    \affiliation{Neutron Scattering Division, Neutron Sciences Directorate, Oak Ridge National Laboratory, Oak Ridge, Tennessee 37831, USA}
    \affiliation{Center for Nanophase Materials Sciences, Physical Science Directorate, Oak Ridge National Laboratory, Oak Ridge, Tennessee 37831, USA}
\author{Mohit Randeria} 
    \affiliation{Department of Physics, The Ohio State University, Columbus, Ohio 43210, USA}
\author{Nandini Trivedi}    
    \affiliation{Department of Physics, The Ohio State University, Columbus, Ohio 43210, USA}
\author{Kenneth S. Burch}    
    \affiliation{Department of Physics, Boston College, Chestnut Hill, Massachusetts 02467, USA}
\author{David C. Bell}    
    \affiliation{Center for Nanoscale Systems, Harvard University, Cambridge, Massachusetts 02138, USA}
    \affiliation{Harvard John A. Paulson School of Engineering and Applied Sciences, Harvard University, Cambridge, Massachusetts 02138, USA}
\author{Frances M. Ross}
    \affiliation{Department of Materials Science and Engineering, Massachusetts Institute of Technology, Cambridge, Massachusetts 02139, USA}
\author{Weida Wu}    
    \affiliation{Department of Physics and Astronomy, Rutgers University, Piscataway, New Jersey 08854, USA}
\author{Don Heiman}    
    \affiliation{Francis Bitter Magnet Laboratory, Plasma Science and Fusion Center, Massachusetts Institute of Technology, Cambridge, Massachusetts 02139, USA}
    \affiliation{Department of Physics, Northeastern University, Boston, Massachusetts 02115, USA}
\author{Valeria Lauter}    
    \affiliation{Neutron Scattering Division, Neutron Sciences Directorate, Oak Ridge National Laboratory, Oak Ridge, Tennessee 37831, USA}
\author{Jagadeesh S. Moodera}    
    \email{moodera@mit.edu}
    \affiliation{Francis Bitter Magnet Laboratory, Plasma Science and Fusion Center, Massachusetts Institute of Technology, Cambridge, Massachusetts 02139, USA}
    \affiliation{Department of Physics, Massachusetts Institute of Technology, Cambridge, Massachusetts 02139, USA}
\author{Hang Chi}
    \email{hang.chi@uottawa.ca}
    %\thanks{These authors contributed equally}
    %\affiliation{Francis Bitter Magnet Laboratory, Plasma Science and Fusion Center, Massachusetts Institute of Technology, Cambridge, Massachusetts 02139, USA}
    %\affiliation{DEVCOM Army Research Laboratory, Adelphi, Maryland 20783, USA}
    \affiliation{Department of Physics, University of Ottawa, Ottawa, Ontario K1N 6N5, Canada}
    \affiliation{School of Electrical Engineering and Computer Science, University of Ottawa, Ottawa, Ontario K1N 6N5, Canada}
    \affiliation{Nexus for Quantum Technologies, University of Ottawa, Ottawa, Ontario K1N 6N5, Canada}
%\date{\today}
%\date{December 20, 2022}
\maketitle

\onecolumngrid

%\tableofcontents

%\clearpage

%\section{\label{sec:level1}Crystal Structure}

%\clearpage

\begin{figure*}%[!thb]
\includegraphics{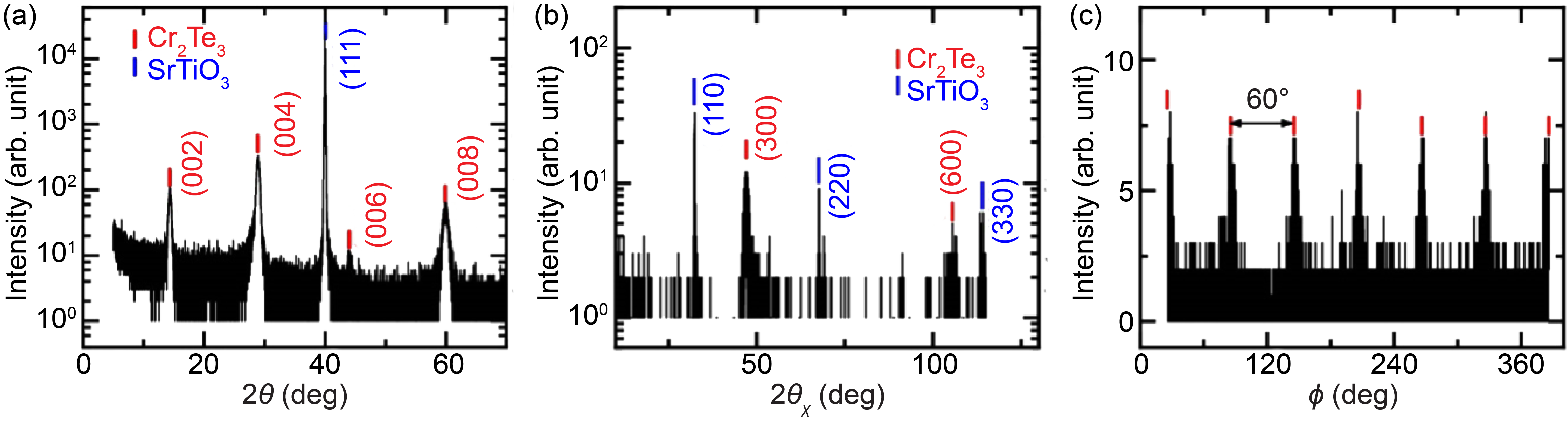}
\mycaption{\label{fig:figs1}
{X-ray diffraction of Cr$_2$Te$_3$ films.} Typical XRD (a) $2\theta$ scan, ({b}) $2\theta_{\chi}$ scan and ({c}) $\phi$ scan of 20 ML Cr$_2$Te$_3$.}
\end{figure*}

\begin{figure*}%[!thb]
\includegraphics{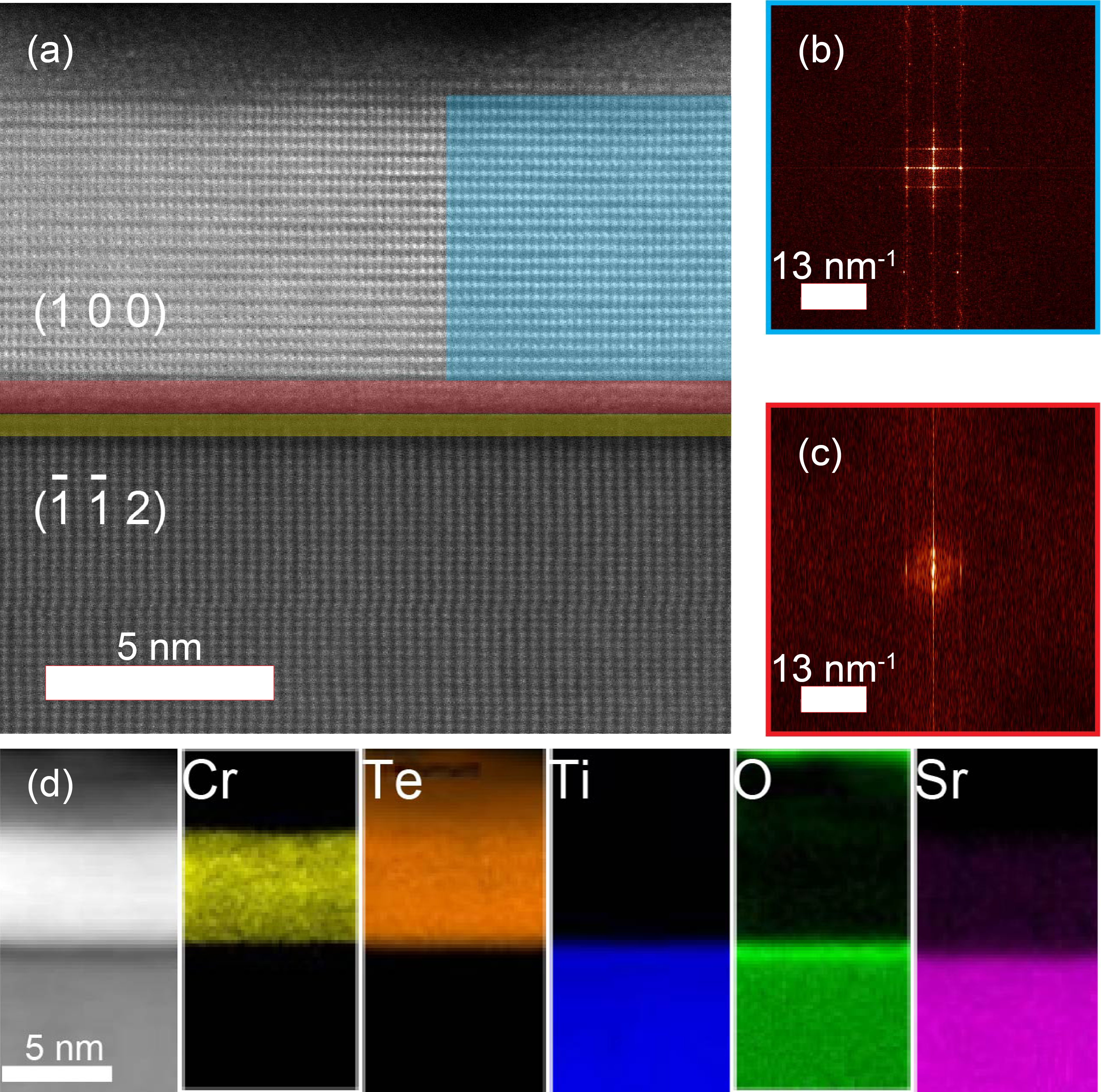}
\mycaption{\label{fig:figs2}
{Scanning transmission electron microscopy of Cr$_2$Te$_3$ films.} ({a}) Large scale HRSTEM HAADF image of 13 ML Cr$_2$Te$_3$ on SrTiO$_3$ for the (100) plane. The first ML of Cr$_2$Te$_3$ on SrTiO$_3$ and the TiO$_x$ top layer of SrTiO$_3$ are indicated by red and yellow. ({b}-{c}) FFT of the ({b}) blue and ({c}) red regions as indicated in ({a}). ({d}) HAADF STEM image (left) and the corresponding EDS elemental mapping.}
\end{figure*}

\begin{figure*}%[htb]
\includegraphics{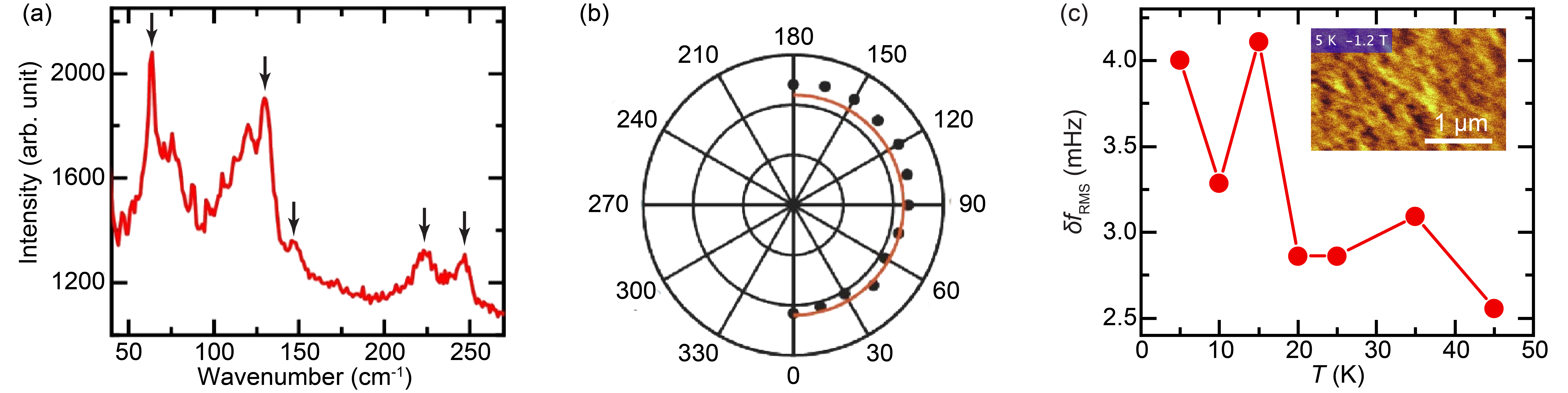}
\mycaption{\label{fig:figs3}
{Raman spectroscopy and magnetic force microscopy of Cr$_2$Te$_3$ films.} ({a}) The Raman spectra of Cr$_2$Te$_3$ with XX polarization at room temperature (arrows indicating 5 possible $A_g$ modes that are independent of the polarization). ({b}) The polar plot of a representative phonon mode of Cr$_2$Te$_3$. ({c}) Temperature dependence of the root-mean-square (RMS) value of the MFM signal. The inset is the MFM image of 1 ML Cr$_2$Te$_3$ scanned at 5~K with a -1.2~T magnetic field applied along the $c$ direction. }
\end{figure*}

\begin{figure*}%[htb]
\includegraphics{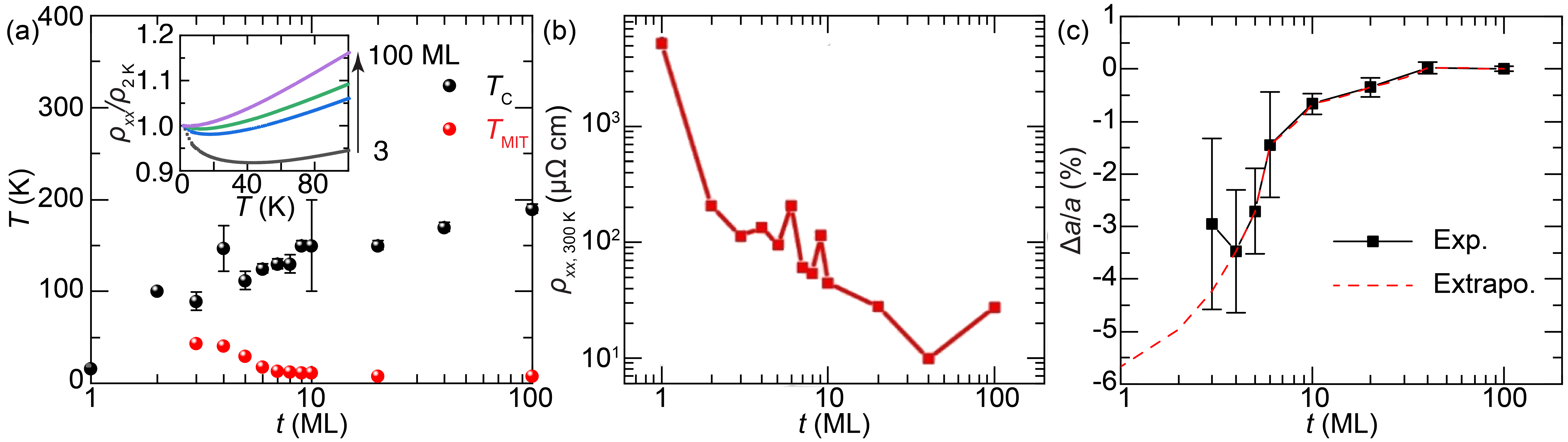}
\mycaption{\label{fig:figs4}
{Thickness dependent physical properties of Cr$_2$Te$_3$ films.} ({a}) Thickness dependence of the Curie temperature $T_{\textrm{C}}$ and the inflection point $T_{\textrm{MIT}}$ in resistivity (i.e., the minima of $\rho_{xx}(T)$ in the inset), displaying a metal-insulator transition-like behavior. ({b}) Layer dependence of the electrical resistivity of Cr$_2$Te$_3$ films at room temperature. ({c}) Layer sensitive relative change in the in-plane lattice parameter $a$.}
\end{figure*}

\begin{figure*}%[htb]
\includegraphics{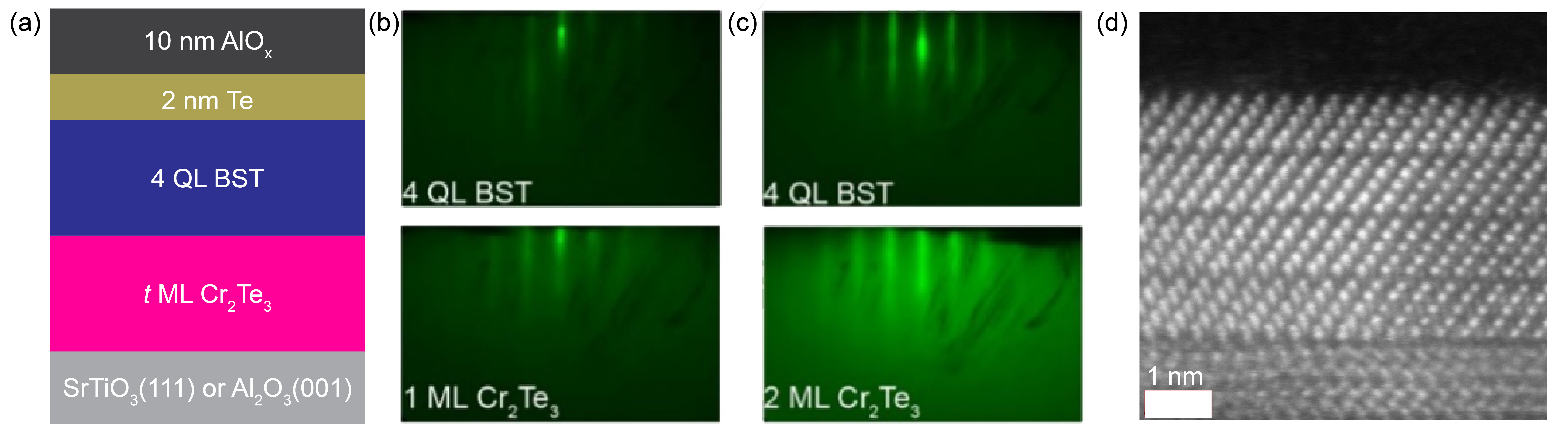}
\mycaption{\label{fig:figs5}
{Few layer Cr$_2$Te$_3$ proximitized with (Bi,Sb)$_2$Te$_3$ topological insulator.} ({a}) Schematic structure of Cr$_2$Te$_3$/(Bi,Sb)$_2$Te$_3$ (BST) heterostructure. ({b}-{c}) RHEED patterns of ({b}) 1 ML Cr$_2$Te$_3$/4 QL BST and ({c}) 2 ML Cr$_2$Te$_3$/4 QL BST along the [100] direction of Cr$_2$Te$_3$. ({d}) STEM image of 2 ML Cr$_2$Te$_3$/4 QL BST.}
\end{figure*}

\begin{figure*}%[htb]
\includegraphics{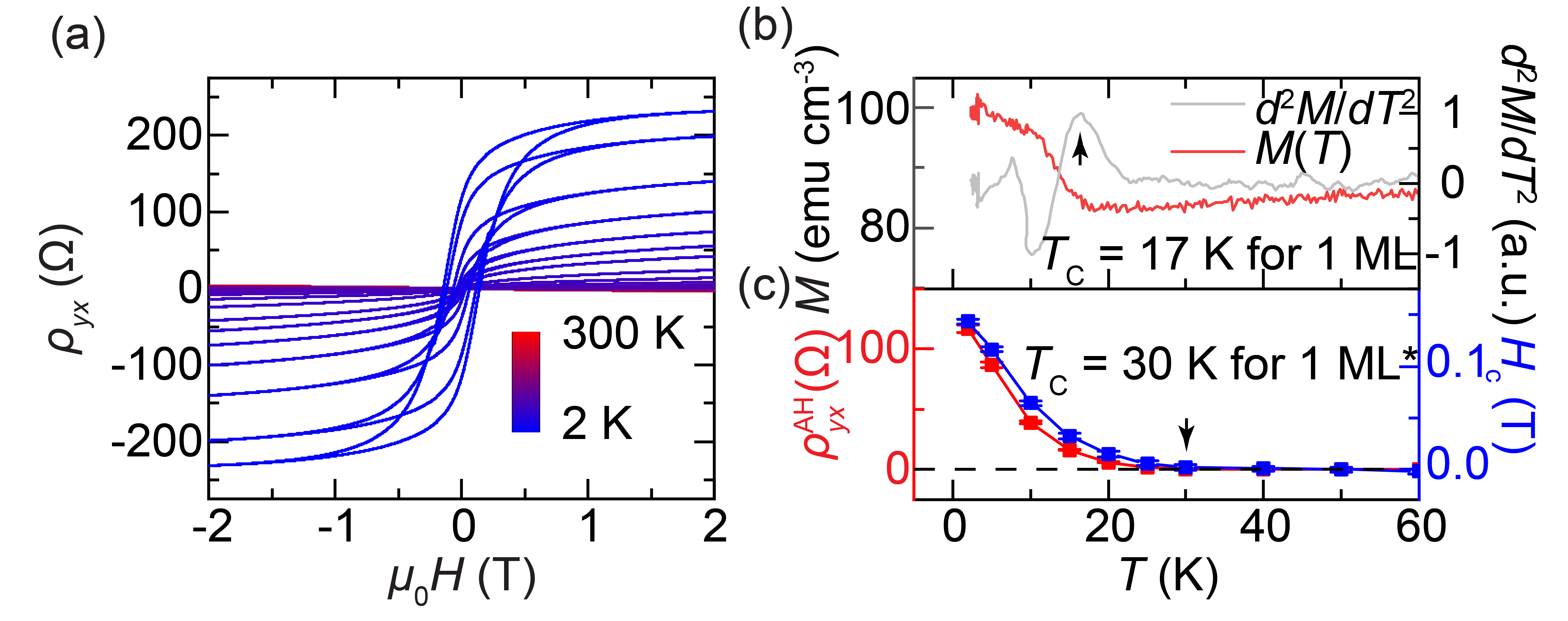}
\mycaption{\label{fig:figs6}
{Topological insulator enhanced ferromagnetism in 1 ML Cr$_2$Te$_3$.} ({a}) Anomalous Hall traces of 1 ML Cr$_2$Te$_3$/4 QL TI (labeled as 1 ML*) at different temperatures from 2 to 300 K. ({b}-{c}) Curie temperature $T_{\rm C}$ estimated from ({b}) the magnetization profile of pristine 1 ML Cr$_2$Te$_3$ film and ({c}) transport data of 1 ML Cr$_2$Te$_3$/4 QL TI, using the remanent anomalous Hall resistance $\rho^{\rm AH}_{yx}(H = 0)$ (red) and the coercive field (blue) extracted from ({a}). Furthermore, field-cool magnetic moment measurements on a similar series of 5 ML chromium telluride vs. heterostructure with 3 QL bismuth telluride samples (data not shown here) also corroborate the enhancement in $T_{\rm C}$ upon TI coupling with $\Delta T_{\rm C} \sim$ 50 K.}
\end{figure*}

\begin{figure*}%[htb]
\includegraphics[width=1\textwidth]{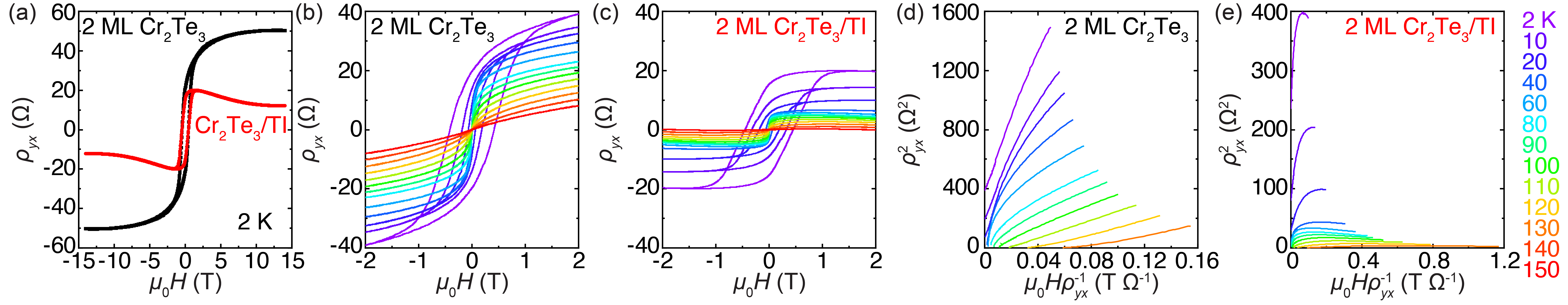}
\mycaption{\label{fig:figs7}
{Topological insulator enhanced ferromagnetism in 2 ML Cr$_2$Te$_3$.} ({a}) Magnetic field dependent anomalous
Hall resistance $\rho_{yx}(H)$ of 2 ML Cr$_2$Te$_3$ (black) and 2 ML Cr$_2$Te$_3$/4 QL TI (red) at $T$ = 2 K, after removing the high field linear-in-H ordinary Hall background up to $\mu_{0}H$ = 14 T. ({b}-{c}) Anomalous Hall and ({d}-{e}) Arrott plot in the magnetic field range of $\pm 2$ T at selected temperatures, for ({b},{d}) 2 ML Cr$_2$Te$_3$ and ({c},{e}) 2 ML Cr$_2$Te$_3$/4 QL TI, respectively.}
\end{figure*}

\begin{figure*}%[htb]
\includegraphics[width=1\textwidth]{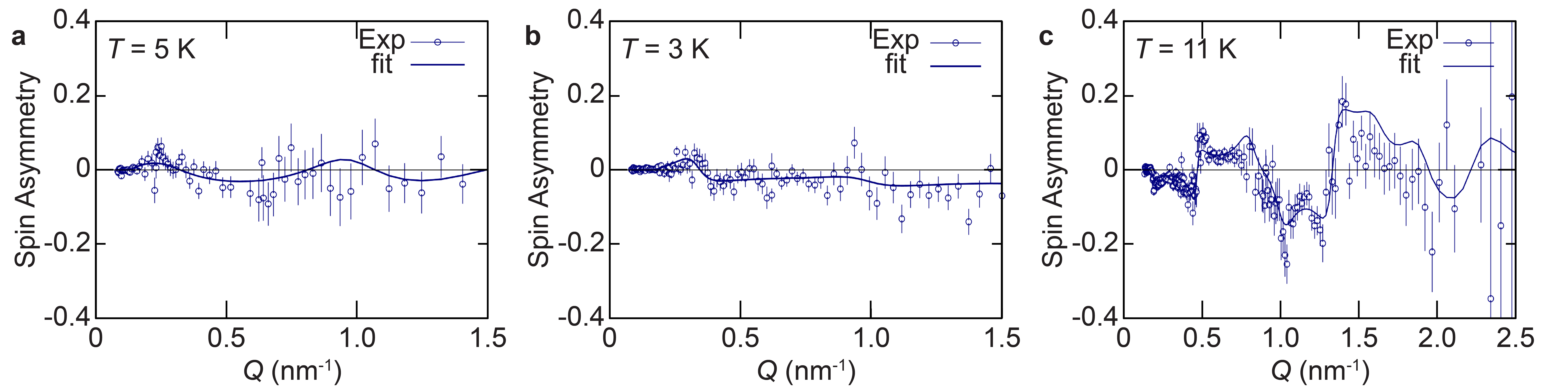}
\mycaption{\label{fig:figs8}
{Polarized neutron reflectometry of Cr$_2$Te$_3$ thin films and their heterostrucutres with (Bi,Sb)$_2$Te$_3$.} PNR spin asymmetry ratio SA = $(R^{+} - R^{-})/(R^{+} + R^{-})$ for pristine Cr$_2$Te$_3$ on ({a}) SrTiO$_3$(111) with in-plane magnetic field $\mu_{0}H$ = 1.0~T at $T$ = 5 K, ({b-c}) on Al$_2$O$_3$(001) with ({b}) $\mu_{0}H$ = 1.0~T and ({c}) $\mu_{0}H$ = 4.8~T at 3 K, ({d}) interfaced (4.8 T, 3 K) and ({e}) sandwiched (1.0 T, 11 K) with (Bi,Sb)$_2$Te$_3$ topological insulator layers, respectively.}
\end{figure*}

\begin{figure*}%[htb]
\includegraphics{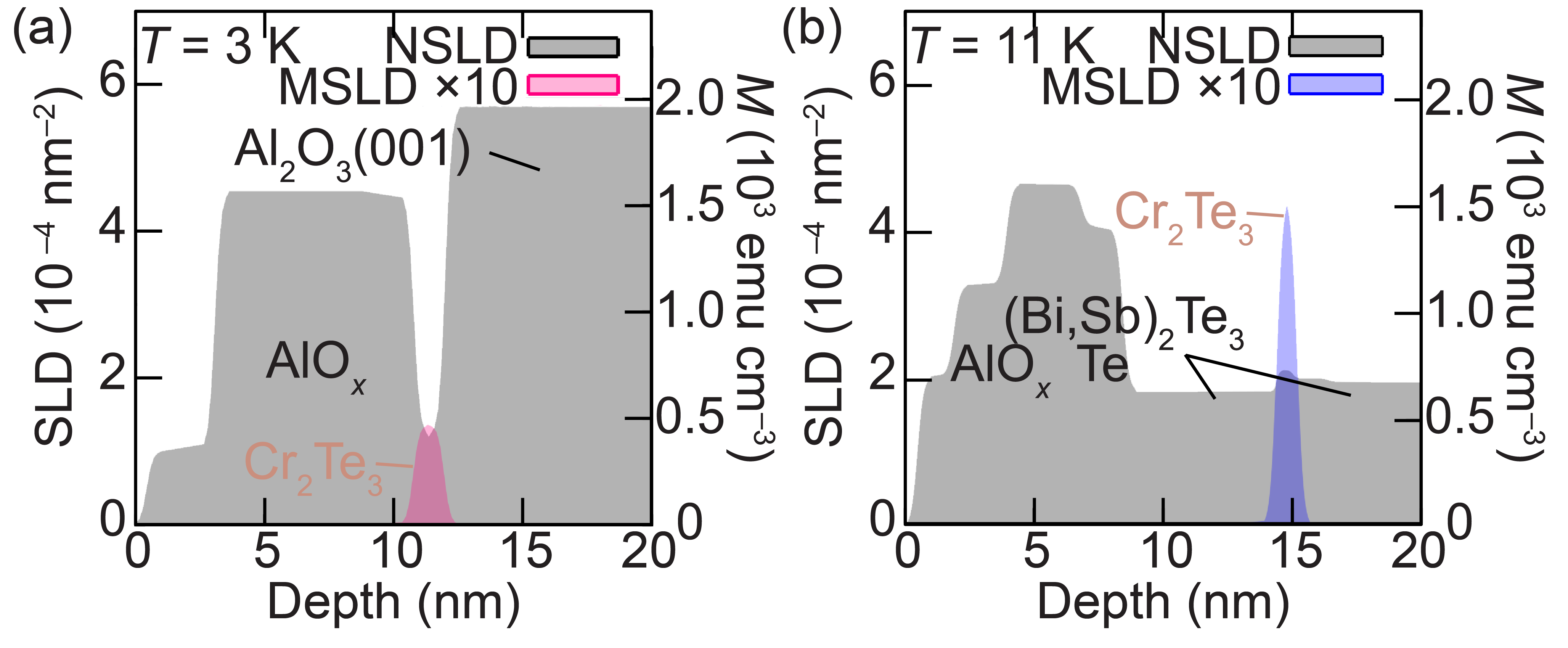}
\mycaption{\label{fig:figs9}
{Topological insulator coupling enhanced magnetism in Cr$_2$Te$_3$.} Depth profiles of nuclear (NSLD) and magnetic (MSLD) scattering length densities (SLD) measured with in-plane magnetic field $\mu_{0}H$ = 1.0 T for (a) Cr$_2$Te$_3$ monolayers at 3~K and (b) sandwiched with (Bi,Sb)$_2$Te$_3$ TI at 11 K, where interface-induced enhancement of magnetism is evident.}
\end{figure*}

% The \nocite command causes all entries in a bibliography to be printed out
% whether or not they are actually referenced in the text. This is appropriate
% for the sample file to show the different styles of references, but authors
% most likely will not want to use it.
%\nocite{*}

%\clearpage
%\def\bibsection{\section*{References}}
%\FloatBarrier

%\begin{minipage}{\textwidth}
%\bibliography{1-SI}
%\end{minipage}

%\bibliography{SI.bib}% Produces the bibliography via BibTeX.